\documentclass[a4paper,11pt]{article}
\usepackage{graphicx} 

\usepackage{graphicx} 
\usepackage{latexsym}	
\usepackage{amsmath}
\usepackage{amsthm}
\usepackage{amssymb}
\usepackage{amsmath}
\usepackage{amsthm}

\usepackage{graphicx}
\def\ba{\begin{eqnarray}}
\def\ea{\end{eqnarray}}

\def\ba{\begin{eqnarray}}
\def\ea{\end{eqnarray}}

\def\be{\begin{equation}}
\def\ee{\end{equation}}

\theoremstyle{plain}

\begin{document}
\title{About  possible measures in Quantum Gravity}
\author{ Osvaldo P. Santill\'an$^1$
\thanks{Electronic  firenzecita@hotmail.com}\\
\textit{\small{$^1$Instituto de Matem\'atica Luis Santaló (IMAS) and CONICET,}}\\
\textit{\small{Ciudad Universitaria, 1428 Buenos Aires, Argentina.}}}
\date{}
\maketitle

\begin{abstract}
There exist several different proposals for a measure in Quantum Gravity theories. Although sometimes being labelled as non covariant, the measure derived in \cite{fradkin}  for GR has the particularity that, in the extremal, the volume divergences proportional to $\delta^4(0)$ cancel out.  The analogous for Quadratic Gravity \cite{stelle1}-\cite{stelle2}  was considered in \cite{buchbinder}-\cite{buchbinder2}.  However, as far as the author's knows, the issue of volume divergences was not considered for this last measure. The present work fills this gap and presents an analysis showing that, in the extremal, these divergences cancel as well. This is up to some subtleties related to superdeterminants.  The possibility of employing non invariant measures may be accepted if the anomaly in the measure is compensated by counter term redefinitions of the model under analysis. This makes difficult to  disprove, at the present times, some choices of measures. Quadratic Gravity\cite{stelle1}-\cite{stelle2}, is known to be renormalizable in flat space, and there are a finite number of counter terms needed in order to renormalize its effective action. However, around a curved space this is not known, and this complicates considerably the analysis. These issues are reviewed in the text, together with an analysis of covariant measures. In particular, it is shown how these measures \cite{botelho}-\cite{mottola2} can be found if one condition in \cite{fradkin} is relaxed.
\end{abstract}
\section{Introduction}

 The choice of a path integral measure is an important point for quantization, in particular for  gravity. In general, it  is desirable to find a realization of this object,  invariant under the symmetries of the model that are considered fundamental. Typical examples are invariance under gauge and coordinate transformations.  However,  the request to work with an invariant measure may be too restrictive. The final goal  is that the $S$ matrix has the same value for different gauges, up to a redefinition of renormalization constants, or that it is generally covariant under general coordinate transformations.  
 
 An equally important problem is how to  impose consistently  two different fundamental symmetries of the model, for instance gauge and coordinate invariance, simultaneously with a single choice of  measure. If this can not be achieved, it may constitute a serious drawback for the quantization of the  model under consideration.

One may insist upon constructing an invariant measure. This type of measure is heavily employed in the Faddeev-Popov-DeWitt \cite{faddeevo}-\cite{dewitt} approach for quantization. A  typical gauge fixed integral is of the form
$$
Z=\int e^{iS[g]}J[g, F(g)]\delta(F(g))d\mu(g),
$$
where $F(g)$ represents the gauge fixing function and the Jacobian $J[g, F(g)]$ in the last expression can be written as
$$
J[g, F(g)]^{-1}=\int \delta(F(g^f))d\mu^h(f),
$$
where $d\mu^h(f)$ is the Haar measure on the gauge or diffeomorphism group orbit, ensuring the invariance $J[g, F(g)]=J[g^f, F(g^f)]$ under diffeomorphisms $f$.  Since the action $S$ is invariant under a diffeomorphism $f$ as well,  this implies that the value of $Z$ depends on the choice of the gauge fixing $F$ through the delta term $\delta(F(g))$, as the measure is invariant $d\mu(g)=d\mu(g^f)$.  This ensures that the "longitudinal modes" are BRST paired with the Faddev-Popov ghosts and decouple from the physical spectrum, leading the BRST invariant fields as the physical modes. The gauge or coordinate invariance of the $S$ matrix was considered in several references such as  \cite{faddeev}-\cite{faddeev2} or \cite{fradkin}, by assuming that the measure is of the above type.

The formal arguments of the previous paragraph however, have to be taken with some care since, in general,  objects such as correlation functions (which are not necessarily gauge invariant neither generally covariant) or effective actions, which are needed for evaluating the $S$ matrix, have to be regularized. One of the well studied  regularization methods is due to Pauli and Villars. In this approach, the regularization  is achieved by adding suitable counter terms in the effective action. The necessity of these  counter terms has the following consequence. A measure may have an anomalous behavior under the action of the set of fundamental  symmetries, that is,
$$
d\mu(g^f)=A(f)d\mu(g),\qquad A(f)\neq 1.
$$
Here "f" is a condensed notation for the symmetry action. The last  behavior may be allowed if the corresponding anomaly term $A(f)$ can be adsorbed by  counter term redefinitions. In this case,  the anomaly may be an illusory effect \cite{nieu}. Therefore, the request for the measure to be invariant may be too restrictive and, depending on the model, some non invariant measures may be allowed as well. In particular, if the $S$ matrix has the desired behavior under the symmetry action, even if  the intermediate steps do not. 

If one is able to find a perfectly invariant  measure under the fundamental symmetries of the model, that is $A(f)=1$, then this is a useful result. If not, it does not necessarily implies that the theory must be abandoned. Furthermore, there may be several different invariant measures equally valid from the theoretical point of view. The allowance of these  measures depends on the model, since the anomaly term has to match the counter terms of the corresponding lagrangian. 

The understanding of this situation took considerable effort, and was recognized in several references such as \cite{lindstrom}-\cite{bardeen}, see also \cite{basti}-\cite{bastibook}.  For instance, for an scalar field $\phi$ in presence of an external gravitational field $g_{\mu\nu}(x)$ some works propose the Liouville measure $D\phi D\pi$ for Hamiltonian path integral \cite{fradkin}. In other words, it is assumed that the Feynman path integral is given by
$$
Z=\int d\mu_L(\pi, \phi, g_{\mu\nu}) e^{i\int d^4x[\pi\dot{\phi}-H(\phi, \pi)]},\qquad \pi=\frac{\delta S}{\delta \dot{\phi}},\qquad d\mu_L(\pi, \phi, g_{\mu\nu})=\prod d\phi \prod d\pi,
$$
the measure corresponds to a unit Jacobian. This measure was motivated by unitarity arguments by Faddeev-Popov \cite{faddeev}-\cite{faddeev2} and by the cancellation of leading one-loop divergences proportional to $\delta^4(0)$ by Fradkin-Vilkovisky \cite{fradkin}.  Toms however, argued that the correct measure is given by  \cite{toms}
$$
d\mu_T(\pi, \phi, g_{\mu\nu})=\prod \frac{1}{\sqrt{g^{00}}} d\phi \prod d\pi.
$$
On the other hand, Hawking  \cite{hawking} and Fujikawa \cite{fujikawa} have imposed a measure in configuration space given by 
$$
d\mu_c( \phi, g_{\mu\nu})=\prod d\widetilde{\phi},\qquad \widetilde{\phi}=g^{\frac{1}{4}}\phi.
$$
The choice for this measure choice is strongly motivated from the physical point of view. It holds by requiring that the Jacobian for the BRST transformation corresponding to coordinate transformations to be a total derivative. In other words, this choice is based on the fundamental requirement of invariance under coordinate transformations. Further choices are found in \cite{unz} and \cite{kaku}.

The problem with the above choices was the following. It is possible to go to a hamiltonian formalism for the Hawking-Fujikawa choice, by introducing the new momentum density $\widetilde{\pi}=(g^{00})^{\frac{1}{2}}g^{-\frac{1}{4}} \pi$, leading to the measure
$$
 d\mu_c(\pi, \phi, g_{\mu\nu})=\prod d\widetilde{\phi} \prod d\widetilde{\pi}.
$$
Since  the Hawking-Fujikawa measure $d\mu_c$ and the Liouville one $d\mu_L$ correspond to different variables, they are not the same. They neither coincide with the Toms ansatz $d\mu_T$.  This suggests a fundamental problem namely, that the measure that ensures unitarity and cancelations of the loop divergences $\delta^ 4(0)$ violates invariance under general coordinate transformations, and the converse of this statement is valid  as well.

Fortunately, these type of arguments are not completely correct. In reference \cite{nieu}  the general measure
$$
 d\mu_c(\pi, \phi, g_{\mu\nu})=\prod d\widetilde{\phi} \prod d\widetilde{\pi}.
$$ was taken into account, where now
$$
\widetilde{\phi}=(g^{00})^{a}g^{b} \phi,\qquad \widetilde{\pi}=(g^{00})^{c}g^{d} \pi,
$$
with $a$, $b$, $c$, and $d$ are arbitrary real numbers. This arbitrary choice implies that $\widetilde{\phi}$ and $\widetilde{\pi}$ are not necessarily canonically conjugated variables.  For a large set of these real numbers, the measure is anomalous under general coordinate transformations. Despite this anomaly, the outcome of reference \cite{nieu} is that the anomalies correspond to different choices of $a$, $b$, $c$, and $d$ are related all by redefinitions of the counterterms  that can be added to the action. This indicates equivalence between the different choices. This means that it is possible, for an scalar field $\phi$, to ensure invariance under coordinate transformations and simultaneously, to ensure unitarity, up to a counter term redefinitions.

This important example shows that  a candidate of a measure can not be discarded without a proper understanding of the counter term structure of the model to be quantized. A non invariant, or anomalous, behavior of the measure is not enough to be rejected, unless it is understood that it induces terms which are not compatible with the set of allowed counter terms of the model. 

Recently, it was recognized the possibility to make Quadratic Gravity, a model which is renormalizable around flat spaces, unitary \cite{salvio}-\cite{salvio2},  by employing a variant of a method of quantization due to Dirac and Pauli \cite{strumia1}-\cite{strumia3}. The measure, which introduces volume divergences of the form $\delta^4(0)$, plays no role in this formulation due to the Veltmann identity $\delta^4(0)=0$ if dimensional regularization is employed. However, perturbations around non trivial curved spaces can be considered as well, for applications such as \cite{gaddam2}-\cite{gaddam3}.  In this context, the choice of measure may play a fundamental role.  There exist different proposals for the measures, some of them \cite{fradkin},  \cite{kaku}, \cite{buchbinder}-\cite{buchbinder2} which contain apparently non covariant factors and generated debate in the literature. The present work is to present a partial description or discussion about these issues.  Recent related references are  \cite{alwis}-\cite{collison}.

The intention of the present work is to study apparently non covariant measures, such as the ones in \cite{buchbinder}-\cite{buchbinder2}, in connection with the cancellation of volume divergences $\delta^4(0)$.
It is well known that these divergences are cancelled in GR \cite{fradkin}, at least at the extremal. However, to the best of the authors knowledge, this analysis has not been done for the measures \cite{buchbinder}-\cite{buchbinder2}. The present work fills this hole, and provides an analysis indicating that this cancellation also holds for Quadratic Gravity. The arguments presented in the text of course, do not imply that this measure is one of the correct ones. But it points out one of the interesting features of dimensional regularization namely, the disappearance of volume divergences. In this sense, the measure found in  \cite{buchbinder}-\cite{buchbinder2} is special.

The present work is organized as follows. In section 2 the possibility of constructing a invariant measure, regardless the model to be quantized, is analyzed. In addition, model dependent measures are also considered, and their proofs of invariance is critically examined. In section 3 it is found that the model dependent measures discussed in section 2, when properly generalized, may lead to volume divergence cancellations, at least in the extremal. This issue is specified for Stelle gravity. Section 4 contains the discussion of the results.

\section{A class of measure candidates}
\subsection{A model independent invariant measure for gravity}

The discussion of the above introduction suggests that  measures that are not completely invariant under the fundamental symmetries of the model may be  allowed. However, it may still be interesting to find such invariant measures. A motivation may be to avoid the analysis of the counter terms of a given model, which can be a hard task. There exist derivations of such invariant measures such as \cite{botelho} or \cite{mottola1}-\cite{mottola5}. In the present section the deduction will follow different lines, much closer to the ones in \cite{fradkin}  since it will make our analysis simpler later on. 

The invariant choice below will be specified for GR or generic gravity theories. Clearly, for those theories, the determination of the structure of the allowed counter terms for quantization along a non flat backgrounds is an epic task.  It would be desirable instead to find a measure 
\begin{equation}\label{meshus}
[Dg^{\mu\nu}]=\prod_x m(g(x)) \prod_x dg^{\alpha\beta},
\end{equation}
 invariant under a general coordinate transformation
$$
y^\mu(x)=x^\mu+\epsilon^\mu(x),\qquad
\overline{g}^{\mu\nu}(y)=\frac{\partial y^\alpha}{\partial x^\mu}\frac{\partial y^\beta}{\partial x^\nu}g^{\alpha\beta}(x),
$$
regardless the model to be quantized. If the parameter $\epsilon(x)$ is infinitesimal, then the last relation becomes
\begin{equation}\label{cosa2}
\overline{g}^{\mu\nu}(y)=g^{\mu\nu}(x)+g^{\mu\alpha}(x)\partial_\alpha \epsilon^\nu(x)+g^{\nu\alpha}(x)\partial_\alpha \epsilon^\mu(x)+O(\epsilon^2).
\end{equation}
The determination of the term $\prod_x m(g(x)) $ can be specified by a careful analysis of the change of the measure under these  general coordinate transformations, by requesting this change to be trivial. This will pose a differential equation for $m(g(x))$. It will be convenient to study this issue in two steps. The first step is to find the change of the measure when passing  from the variable $\overline{g}^{\mu\nu}(y)$ to $\overline{g}^{\mu\nu}(y(x))$.  The second step requires to do the same, but passing from $\overline{g}^{\mu\nu}(y(x))$ to $g^{\mu\nu}(x)$.  The first step implies passing from
\begin{equation}\label{empezar}
 \prod_y m(\overline{g}(y)) \prod_y d\overline{g}^{\mu\nu}(y)=M_2  \prod_x m(\overline{g}(y(x))) C_2\prod_x d\overline{g}^{\mu\nu}(y(x)),
\end{equation}
the quantity $C_2$ is the result from the change of coordinates in the differentials and $M_2$ the change of the measure term $\prod_x m(g(x)) $. These two quantities have to be calculated  explicitly. After this step is achieved,  one must go from 
$$
 \prod_x m(\overline{g}(y(x))) \prod_x d\overline{g}^{\mu\nu}(y(x))=M_1\prod_x m(g(x)) C_1\prod_xdg^{\mu\nu}(x).
$$
 The last relation defines the quantities $M_1$ and $C_1$, which must also be determined explicitly. Altogether, the last two formulas imply that
 $$
 \prod_y m(\overline{g}(y))=M_1 M_2\prod_x m(g(x)),\qquad \prod_y d\overline{g}^{\mu\nu}(y)=C_1 C_2\prod_x dg^{\mu\nu}(x).
 $$
 It is clear that the factor $M_1 M_2 C_1 C_2$ describes the change of the measure under general coordinate transformations. The measure will be invariant if $$M_1 M_2 C_1 C_2=1.$$
This requirement will impose a differential equation for  $m(g(x))$, thus defining the corresponding measure.

After presenting the outline of the general procedure given above, it is now necessary to determine all the factors $M_1$, $M_2$, $C_1$ and $C_2$ in closed form, or ar least, their products.  The product factor $M_1M_2$ is very easily calculated from \eqref{cosa2}. This formula leads to $d^4y=(1+\partial_\mu \epsilon^\mu)d^4x$ and to
 $$
 \prod_y m(\overline{g}(y))=\exp\{\delta^4(0) \int \log(m(\overline{g}(y)))d^4y\}=\exp\{\delta^4(0) \int \log(m(g(x)))d^4x\}
 $$
 $$
\times \exp\{\delta^4(0) \int \log(m(g(x)))\partial_\mu \epsilon^\mu d^4x\} \exp\{\delta^4(0) \int \frac{1}{m(g(x))}\frac{\partial m(g)}{\partial g^{\alpha\beta}}(g^{\alpha\nu }\partial_\nu \epsilon^\beta+g^{\beta\nu}\partial_\nu \epsilon^\alpha) d^4x\}.
$$ 
$$
=\exp\{\delta^4(0) \int \bigg[\log(M(g(x)))\partial_\mu \epsilon^\mu+\frac{1}{m(g(x))}\frac{\partial m(g)}{\partial g^{\alpha\beta}}(g^{\alpha\nu }\partial_\nu \epsilon^\beta+g^{\beta\nu}\partial_\nu \epsilon^\alpha)\bigg] d^4x\}\prod_x m(g(x)).
$$
In other words
$$
M_1 M_2=\exp\{\delta^4(0) \int \bigg[\log(m(g(x)))\partial_\mu \epsilon^\mu+\frac{1}{m(g(x))}\frac{\partial m(g)}{\partial g^{\alpha\beta}}(g^{\alpha\nu }\partial_\nu \epsilon^\beta+g^{\beta\nu}\partial_\nu \epsilon^\alpha)\bigg] d^4x\}.
 $$
 The factor $C_1$ is also not difficult to be calculated. In fact, from its definition
 $$
 \prod_x d\overline{g}_{\mu\nu}(y(x))= C_1\prod_x dg_{\mu\nu}(x),
 $$
 the Jacobian that follows from \eqref{cosa2} is
\begin{equation}\label{terminar}
C_1=\det\bigg[\frac{\delta \overline{g}^{\mu\nu}(y(x))}{\delta g^{\alpha\beta}(x)}\bigg]=1+5\delta^4(0)\int \partial_\mu \epsilon^\mu d^4x\simeq\exp(5\delta^4(0)\int \partial_\mu \epsilon^\mu d^4x).
 \end{equation}
 It is customary to state that the integral of a total divergence vanishes. This will lead to the conclusion that $C_1$=1. However, at this point, it may be convenient to keep the Jacobian in this form. The reason is that to disregard the total divergence  may be dangerous. For instance,  in the Fujikawa method applied to gauge fields there are terms such that $\delta^4(0) \text{Tr} 
 \gamma_5$ which are naively zero due to the traceless property of $\gamma_5$. However, when the divergent delta factor $\delta^4(0)$ is regularized in gauge invariant form, this term becomes the well known ABJ anomaly proportional to $F_{\mu\nu}\widetilde{F}^{\mu\nu}$. Therefore, it is convenient to keep the term $5\delta^4(0)\int \partial_\mu \epsilon^\mu d^4x$. If its presence is irrelevant, then this will be clear as a process of an explicit calculation, to be described below. 

At this point, no regularization of $\delta^4(0)\int \partial_\mu \epsilon^\mu d^4x$ is presented. The reason is that if the terms of this form are cancel each other out before regularization, then it is not needed to specify it. This is precisely one of the conditions to be requested below.

The last step is to calculate the factor $C_2$ which arises when passing  from coordinates $y^\mu$ to $x^\mu$ without changing the frame
$$
\prod_y d\overline{g}^{\alpha\beta}(y)=C_2\prod_x d\overline{g}^{\alpha\beta}(y(x)).
$$
In other words, this factor is related to the description of the metric in the frame $y^\mu$ but whose components are parametrized in terms of $x^\mu$. An approach for the calculation of this factor may be to expand the metric components in a Fourier basis
$$
\overline{g}^{\alpha\beta}(y)=\frac{1}{\sqrt{2\pi}}\int e^{ipy}\overline{g}^{\alpha\beta}(p)  d^4p,\qquad \overline{g}^{\alpha\beta}(p)=\frac{1}{\sqrt{2\pi}}\int e^{-ipy}\overline{g}^{\alpha\beta}(y)  d^4y,
$$and to employ these coefficients as the new integration variables.  The change to these variables is given by
$$
\prod_y d\overline{g}^{\alpha\beta}(y)=\det(\frac{e^{ipy}}{\sqrt{2\pi}})\prod_p d\overline{g}^{\alpha\beta}(p).
$$
The Fourier matrix transformation $\frac{e^{ipy}}{\sqrt{2\pi}}$ is unitary, thus the determinant in the last expression may be set to one. Under the coordinate change $y^\mu=x^\mu+\epsilon^\mu(x)$ one would like to compare the  last coefficients with the ones related to $x^\mu$, namely, with the coefficients
$$
\widehat{g}^{\alpha\beta}(p)=\frac{1}{\sqrt{2\pi}}\int e^{-ipx}\overline{g}^{\alpha\beta}(y(x))  d^4x\simeq
\overline{g}^{\alpha\beta}(p)+\frac{1}{\sqrt{2\pi}}\int e^{-ipx}\epsilon^\mu(x)\partial_\mu \overline{g}^{\alpha\beta}(x) d^4x.
$$
A simple inspection of the last formula shows that the last  are the Fourier coefficients correspond to the deformed metric 
$$
\overline{g}^{\mu\nu}_\epsilon(x)=\overline{g}^{\mu\nu}(x)+\epsilon^\alpha(x)\partial_\alpha \overline{g}^{\mu\nu}(x).
$$
From here it is concluded, by going again to the spatial coordinates, that
$$
\prod_y d\overline{g}^{\alpha\beta}(y)=\prod_x d\overline{g}_\epsilon^{\alpha\beta}(x).
$$
The final step is to pass from $\prod_x d\overline{g}_\epsilon^{\alpha\beta}(x)$ to $\prod_x d\overline{g}^{\alpha\beta}(x)$. This requires the calculation of the Jacobian
$$
J=\det \frac{\delta \overline{g}_\epsilon^{\mu\nu}(x)}{\delta \overline{g}^{\mu\nu}(x)}=\det(1+\epsilon^\mu(x)\partial_\mu).
$$
The last determinant is
$$
\det(1+\epsilon^\mu(x)\partial_\mu)=\exp(\text{Tr}\log(1+\epsilon^\mu\partial_\mu))\simeq \exp(\text{Tr}(\epsilon^\mu\partial_\mu))
$$
$$
=\exp(\sum_p \int e^{-ipx}\epsilon^\mu\partial_\mu e^{ipx}d^4x)=\exp(i\sum_pp _\mu \int \epsilon^\mu d^4x)=1
$$
since the sum (or integral) of $p_\mu$ in all the range of integration is zero, as $p^\mu$ is an odd function. Therefore
$$
\prod_y d\overline{g}^{\alpha\beta}(y)=\prod_x d\overline{g}_\epsilon^{\alpha\beta}(x)=\prod_x d\overline{g}^{\alpha\beta}(x),
$$
in other words $C_2=1$. This is precisely the point of discussion with other references, which calculate that $C_2$is instead non trivial \cite{fradkin}, \cite{buchbinder}-\cite{buchbinder2}, and this point will be studied in the following sections.

By collecting all the above results
$$
 \prod_x m(\overline{g}(y) \prod_y d\overline{\phi}_a(y)=M_1M_2C_2 \prod_x m(g(x)) C_1\prod_x d\phi_a(x),
$$
$$
=\exp\{\delta^4(0) \int \bigg[\log(m(g(x)))\partial_\mu \epsilon^\mu+\frac{1}{m(g(x))}\frac{\partial m(g)}{\partial g^{\alpha\beta}}(g^{\alpha\nu }\partial_\nu \epsilon^\beta+g^{\beta\nu}\partial_\nu \epsilon^\alpha)\bigg] d^4x\}
$$
\begin{equation}\label{factores}
\exp(5\delta^4(0)\int \partial_\mu \epsilon^\mu d^4x)\prod_x m(g(x))\prod_x d\phi_a(x).
\end{equation}
The final task is to fix $m(g(x))$ in such a way  that the factors \eqref{factores} conspire to give a unit value. 
This condition leads to the following differential equation for $m(g(x))$
$$
\log(m(g(x)))\partial_\mu \epsilon^\mu+\frac{1}{m(g(x))}\frac{\partial m(g)}{\partial g^{\alpha\beta}}(g^{\alpha\nu }\partial_\nu \epsilon^\beta+g^{\beta\nu}\partial_\nu \epsilon^\alpha)-5\partial_\mu \epsilon^\mu=0.
$$
Since $\delta g=g g_{\mu\nu}\delta g^{\mu\nu}$ it is natural to propose that $M(g)$ depends only on $g$. Then
$$
\bigg[\log(m(g))+2 g\frac{M^{\prime}(g)}{m(g)}-5\bigg]\partial_\mu \epsilon^\mu=0.
$$
The terms in brackets vanishes if
$$
\int \frac{dg}{g}=\int \frac{2dm}{m(\log m-5)}
$$
leading to
$$
m(g)=\exp\{\frac{\alpha}{\sqrt{-g}}+5\}.
$$
This expression does not contain factors proportional to $g^{00}$. In fact, if $\alpha$ goes to zero, the resulting factor becomes constant, thus the simple flat measure
\begin{equation}\label{mia}
d\mu_f(g)=\prod_x dg^{\alpha\beta},
\end{equation}
is a valid choice as well. The factor $5$ is due to the total divergence term, which plays no role. However, this has been checked here explicitly, no assumptions have been made to discard this term, it is the result of a concrete calculation. Needless to say, this is the simplest measure that one may imagine.

\subsection{Model dependent choices and the issue of  $g^{00}$ factors.}

The above result may lead to apparent issues with some existing literature, and it is a good point to clarify them, at least partially. 

For quantization of GR, the following measure
\begin{equation}\label{relato}
[Dg_{\mu\nu}]_s=\prod_{x} dg_{\mu\nu}(x) g^{00}(x) |g(x)|^{-\frac{3}{2}}.
\end{equation}
has been worked out in \cite{fradkin}. For Stelle gravity \cite{stelle1}-\cite{stelle2} instead, the following choice has been proposed \cite{buchbinder}
$$
[Dg_{\mu\nu}]_s=\prod_{x} dg_{\mu\nu}(x) (g^{00}(x))^4 |g(x)|^{-\frac{3}{2}}.
$$
This measure corresponds to a generic choice of parameters of the model. For certain particular combinations, this measure may degenerate into something functionally different \cite{buchbinder}. Further choices for higher dimensional gravity models can be found in \cite{buchbinder2}.
All these measures  correspond to the Liouville measure $d\mu_L=\prod d\phi d\pi$ with $\phi$ and $\pi$ are canonically conjugated momenta, after integrating the momenta. The results are model dependent and of course, are not the same than the one found above in  \eqref{mia}.

In fact, a curious feature is that the measures $[Dg_{\mu\nu}]$ given above are different for GR and for Stelle gravity, even though GR can be obtained as a limit from Stelle gravity. The Stelle lagrangian is 
\begin{equation}\label{geld}
L_2=-\bigg[-\kappa^{-2}R+\frac{\alpha}{2} R_{\mu\nu}R^{\mu\nu}+\beta R^2\bigg]\sqrt{-g},
\end{equation}
where the curvature is given as
$$
R^\beta_{\alpha \mu\nu}=\partial_\mu \Gamma_{\nu\alpha}^\beta-\partial_\nu \Gamma_{\mu\alpha}^\beta+\Gamma_{\mu\gamma}^\beta\Gamma_{\nu\alpha}^\gamma-\Gamma_{\nu\gamma}^\beta\Gamma_{\mu\alpha}^\gamma,
$$
while $R_{\alpha\nu}=R^\beta_{\alpha \beta\nu}$ is the Ricci tensor and $R=g^{\mu\nu}R_{\mu\nu}$ the scalar curvature.  The parameters $\alpha$
and $\beta$ are expected to be small corrections to GR. If they tend to zero, GR is recovered. The measure function $M(g(x))$ instead changes abruptly its functional form when $\alpha$ and $\beta$ tend to zero. This is  perhaps a curious feature, but not necessarily wrong. In fact, in this limit, renormalizability is completely spoiled.  Furthermore, the Stelle equations of motion, which are of fourth order, degenerate into a second order system. Therefore, the measure may somehow degenerate in something functionally different when the GR limit is taken. 

There are proofs of the invariance under diffeomorphisms of the measures presented in the previous paragraph. It is convenient to review these arguments in certain detail here. In particular, it  is worthy to review the main reasoning of reference \cite{fradkin} since, if there is a flaw, it is very subtle and to detect it will point to a deep question.   

Indeed, one of the most debated points of the GR  measure \eqref{relato} is the presence of the non covariant factors proportional to $g^{00}$. These terms arise as follows.  The quantization method  employed in \cite{fradkin} starts with the Hamiltonian formulation of the model  $H(p_i,q_i)$ and assumes that the path integral is given by
$$
Z=\int [Dp_i][Dq_i] \exp \{i \int [p^i\dot{q}_i-H(p^i,q_i)]d^4x\}.
$$
Here $p^i$ and $q_i$ are canonical variables and the measure choice is simply
$$
[Dp_i][Dq_i] =V\prod_x dq_i(x) \prod_x dp^i(x). \qquad V=1.
$$
In other words,  the infinite dimensional volume term factor is unity.  This is known as the Liouville measure. This quantization is possible under certain non degeneracy assumptions. If instead constraints appear, there are different methods to go to the Hamiltonian formalism with a suitable choice of reduced canonical variables. An example is Dirac quantization, a technique employed heavily in \cite{buchbinder}-\cite{buchbinder2}. 

In any case, assume that non degeneracy takes place. The reason for which the Liouville measure is attractive is that the action of a symmetry on the generic fields can be described by a canonical transformation of the form
$$\delta q^i=\{q^i, Q\}\qquad \delta p_i=\{p_i, Q\},$$ where  $Q$ is the Noether charge that induces the symmetry action. The Liouville action is expected to be invariant under canonical transformations, in particular, those induced by $Q$. This indicates that the Liouville measure is a choice which respects symmetry transformations in general. This point will be critically reconsidered below.

After choosing the Liouville measure, the next step is to pass to the  lagrangian  path integral formalism. This is obtained by integrating the momenta $p_i$, leading to an expression of the form
$$
Z=\int e^{iS(\phi_a(x))} \prod_x m(g(x)) \prod_x d\phi_a(x),
$$
and the desired lagrangian measure can be read from this version, namely
$$
 [D\phi_a(x)]=\prod_x m(g(x)) \prod_x d\phi_a(x).
$$
The measure is of course model dependent, as the result of the integration on $p_i$ depends on the lagrangian in question.

All the above arguments are valid if the action of symmetries $\delta q^i=\{q^i, Q\}$, $\delta p_i=\{p_i, Q\}$ is not anomalous. These symmetry actions in fact, may induce a divergent Jacobian $J$ whose regularization may break them down, unless it can be adsorbed by counter terms redefinition. The methods \cite{fradkin} given above for obtaining the GR measure were elaborated before 1973, and  the role of the divergent Jacobians for studying anomalies was not understood at those times. Therefore it is a good point to analyze these measures again, with a more modern point of view, by keeping an eye at this anomaly issue.

After considering the above general procedure outlined above for obtaining the measure, the authors \cite{fradkin} apply it to lagrangians of the form
\begin{equation}\label{cosa}
L=D^{ab\mu\nu}(\phi_a) \partial_\mu \phi_a \partial_\nu \phi_b +E^{a\mu}(\phi_a) \partial_\mu \phi_a+F(\phi_a).
\end{equation}
The fields $\phi_a$ are generic fields. A key point is that GR in Dirac coordinates can be written in  this form. In such theory,  $\phi_a(x)$ may be related to  the metric field $g_{\alpha\beta}(x)$ or any other dynamical field of the model, the discussion at this point is quite generic.   Under specific non degeneracy assumptions about the quantities $D^{ab\mu\nu}(\phi_a) $,  the Hamiltonian version of these theories can be derived straightforwardly.  The method  starts with the Hamiltonian path integral of the theories \eqref{cosa} with the Liouville choice of measure
$$
Z=\int e^{i\int [\dot{\phi}_a \pi_a-H(\pi_a, \phi_a)]d^4x } \prod_x d\pi_a(x) \prod_x d\phi_a(x).
$$
After integration in $\pi_a(x)$, the resulting path integral in lagrangian form is
$$
Z=\int e^{iS(\phi_a(x))} \prod_x\det\bigg[\frac{\delta^2 L}{\delta (\partial_0\phi_a)\delta (\partial_0\phi_b)}\bigg] \prod_x d\phi_a(x).
$$
In principle, this identifies the measure as
\begin{equation}\label{meshu}
\prod_x m(g(x)) \prod_x d\phi_a(x)= \prod_x\det\bigg[\frac{\delta^2 L}{\delta (\partial_0\phi_a)\delta (\partial_0\phi_b)}\bigg] \prod_x d\phi_a(x),
\end{equation}
which, as anticipated, is lagrangian dependent. 

One may wonder why the Liouville measure described above is the most privileged one, specially after the discovery of anomalies \cite{fujikawas}. However, there is another attractive feature about the Liouville choice. This will be discussed  here briefly,  details will be given in the next sections.  First, note that this measure can be written as
$$
\prod_x m(g(x)) \prod_x d\phi_a(x)= \prod_x\exp\bigg\{\delta^4(0)\int \text{Tr}\bigg[\frac{\delta^2 L}{\delta (\partial_0\phi_a)\delta (\partial_0\phi_b)}\bigg]d^4x\bigg\} \prod_x d\phi_a(x).
$$
The presence of the volume divergences $\delta^4(0)$ terms may be problematic. However, the second order variation of the action 
$$
S=S_0+\int\frac{\delta^2S}{\delta \phi_a \delta \phi_b}\delta\phi_a\delta\phi_bd^4x+..
$$
around classical solutions $\phi_{a0}$ give rise to a divergent term related to the 
$$
\det(\frac{\delta^2S}{\delta \phi_a \delta \phi_b}).
$$
There are therefore two divergent term $\delta^4(0)$ which, as argued in \cite{fradkin} cancel each other. Another choices will lead to a propagating $\delta^4(0)$ which  may lead to several complications in the Feynman diagrams. This  feature is a fundamental motivation for keeping the Liouville choice, which leads to \eqref{meshu}.  

In fact, below, more will be said about this fact, but the analysis will be focused on Quadratic Gravity instead of GR. 

Even taking into account the above motivation,  it is still needed is to clarify the behavior of the last measure under general coordinate changes.  This analysis  is pretty similar to the one presented in the  previous section, but  with a single difference. As in the previous section,  the strategy is to go first from $\overline{g}^{\mu\nu}(y)$ to $\overline{g}^{\mu\nu}(y(x))$  and then from $\overline{g}^{\mu\nu}(y(x))$ to $g^{\mu\nu}(x)$. This procedure was employed in  \eqref{empezar}-\eqref{terminar}. However, there will be a discrepancy related to the factor $C_2$ which, in  \eqref{empezar}-\eqref{terminar}  was determined to be $C_2=1$\footnote{It is interesting to note that in a previous work \cite{fradkin2}, an author of \cite{fradkin}  considered the possibility to have $C_2=1$. The work \cite{fradkin} then regrets this choice, and employs another route,  described above and in \cite{branchina}.}. A recent pedagogical calculation of the above steps was presented in \cite{branchina}, for this reason the description below will be considerably more brief.

The main observation for the calculation of $C_2$ in \cite{fradkin} is that the path integral $Z$ in Hamiltonian form can not change its value under a redefinition of the variables $\phi_a(x)\to \overline{\phi}_a(x)$ if the momentum $\pi_a$ is redefined accordingly. In other words, the value of $Z$ is unchanged under canonical transformations.  Assuming this, consider again the three choices for the generalized coordinates done above namely, the metric $g^{\mu\nu}(x)$ in a coordinate system $x^\mu$, the same metric $\overline{g}^{\mu\nu}(y)$ in another coordinate system $y^\mu$ and the metric in the frame $y^\mu$ but described in terms of the coordinates $x^\mu$, that is, $\overline{g}^{\mu\nu}(y(x))$ .  The same considerations hold for the ghost fields. Any of the above three choices gives  three different  conjugate momentum densities $\pi^i_a$ with $i=1,2,3$.  At  the classical level, all these choices are represented by a  lagrangian of the form
\begin{equation}\label{cosas}
L=\widetilde{D}^{ab\mu\nu}(\phi_a) \partial_\mu \phi_a \partial_\nu \phi_b +\widetilde{E}^{a\mu}(\phi_a) \partial_\mu \phi_a+\widetilde{F}(\phi_a),
\end{equation}
which coincides with the type described in  \eqref{cosa}. The value of $Z$, under the above assumptions, should be the same for all these choices.  This fact allows to compare the measure obtained by use of the mixed coordinate $\overline{g}^{\mu\nu}(y(x))$ and its conjugated momentum with the one with the pure coordinates $g^{\mu\nu}(x)$ or $\overline{g}^{\mu\nu}(y)$. This comparison, together with the procedure analogous to \eqref{empezar}-\eqref{terminar}, allows to conclude that the resulting factor is \cite{fradkin}
$$
C_2=\frac{ \prod_y m(\overline{g}(y)) }{M_2\prod_x m(\overline{g}(y(x)))}.
$$
The authors \cite{fradkin} pursue this calculation explicitly, and find that  $M_1M_2C_1 C_2=1$ with the above calculated $C_2$ factor. In these terms, they conclude that the measure 
$$
\prod_x m(g(x)) \prod_x d\phi_a(x),
$$
is invariant under general coordinate transformations.  The resulting measure is then
$$
[Dg_{\mu\nu}]_s=\prod_{x} dg_{\mu\nu}(x) g^{00}(x)|g(x)|^{\frac{3}{2}}.
$$
as anticipated in \eqref{relato}. 

The deduction made in \cite{fradkin}  of course, relies on that the canonical transformations just described are not anomalous. They may be instead.  If this is the case, then the statement that  for all the three choices of variables made above, the resulting lagrangian is of the form \eqref{cosa}, may be a source of debate.  On the other hand, as discussed at the beginning, a discrepancy between two measures or the presence of an anomaly does not imply that one of them has to be discarded. Two measures are equivalent if their difference imply a counter term redefinition. And the point is that, in GR, there are infinite counter terms since the theory is not renormalizable. This makes this analysis a hard topic and, in authors opinion, this is the source of the large debate about the validity of the measure of \cite{fradkin}.

\subsection{Comparison with further literature}

Before to proceed to the study of higher order theories, in particular Quadratic Gravity \cite{stelle}-\cite{stelle2}, it is convenient to make a further comparison with literature. 

Several works about  invariant measures can be found in the literature. References such as \cite{botelho}  introduce a Riemann structure into the functional manifold of the metric components, by requiring  compatibility with the invariance group of the theory \cite{dewitt}. An interesting feature of these approaches is that there is no needed to introduce the ad hoc insertion of the Faddeev-Popov unity resolution into the path-lntegral measure in order to extract the gauge orbit volume, since the calculation is performed in a purely geometric way. It is neither needed to go to the Hamiltonian formulation of the path integral. This formalism was developed further in \cite{mottola1}-\cite{mottola5}, in particular, by paying attention to the  analogy with the Polyakov formalism for quantizing strings. In four dimensions, the results of those references reduce to the formula \eqref{mia}  found above.

A further proposal for the measure, related to the Liouville formalism, was introduced in  \cite{unz}. The method employed in that reference starts with the full gauge fixed action with ghosts and auxiliary fields in lagrangian form  After that, it goes to the hamiltonian formalism for the full set of fields, employ the Liouville measure and find the measure after integrating the momenta. This procedure mimics the ones outlined in \cite{fradkin}, the main difference is that now it involves the ghost fields as well. It is pointed out in \cite{unz} that, in  presence of Grassmann variables such as ghost ones,  the measure
$$
\prod_x M(g(x))\prod_x d\phi_a(x)=\prod_x\det\bigg[\frac{\delta^2 L}{\delta (\partial_0\phi_a)\delta (\partial_0\phi_b)}\bigg] \prod_x d\phi_a(x).
$$
has to be replaced by
$$
\prod_x M(g(x))\prod_x d\phi_a(x)=\prod_x S\det\bigg[\frac{\delta^2 L}{\delta (\partial_0\phi_a)\delta (\partial_0\phi_b)}\bigg] \prod_x d\phi_a(x).
$$
where the subscript "S" indicates a super determinant, a generalization of the ordinary determinant for this case, whose definition is given in \cite{dewito}.  It is not mandatory for the purposes to work  out its explicit expression. The main point of \cite{unz} is that there can be corrections to the measure, which become important at the Planck scale. However,  as far as i understand, these corrections do not exclude the debated $g^{00}$ factor.

Another possibility is to employ, in the hamiltonian formalism, a general measure instead of Liouville one. As the authors \cite{fradkin} emphasize, this leads of propagation of terms in the Feynman diagrams proportional to $\delta^4(0)$. While this is true, it may be the case that a regularization exists which takes care of these terms. For flat spaces, dimensional regularization sends these terms to zero. In curved space, the question seems more involved, however the possibility of such regularization is not excluded. 

The measure \cite{fradkin} therefore may be valid. If it is not, the most reasonable possibility for a flaw is, in authors opinion, the possibility of an anomaly when changing the variables from the pure metric $g_{\mu\nu}(x)$ to the mixed one $g_{\mu\nu}(x(y))$. A further possibility is of course that the measure presented in \cite{fradkin}  is  correct.

\section{About the cancellation of $\delta^4(0)$ terms in some quartic theories}

Having studied the  measure for GR, a further step may be to consider the Stelle gravity \cite{stelle1}'\cite{stelle2}, which is  a quantum gravity model, renormalizable when expanded around flat space, and can be made unitary \cite{salvio}-\cite{salvio2} by employing a Dirac-Pauli like quantization \cite{strumia1}-\cite{strumia3}. The Stelle gravity  is described by the lagrangian \eqref{geld} which is of second order in time derivatives of the metric, therefore the corresponding equations of motion are of fourth order. In the following, the quantization of this model with the  Liouville measure will be considered.

\subsection{A preliminary higher derivative model}
Consider a second order lagrangian density for a set of fields $\phi^a$ and  $\eta_a$ of the following form
$$
L=A^{ab}(\phi_a,\eta_a)\ddot{\phi}_a \ddot{\phi}_b+B^{a}(\phi_a,\dot{\phi}_a,\eta^a) \ddot{\phi}_b+C(\phi_a,\dot{\phi}_a, \eta_a)+D^{ab}(\phi_a, \dot{\phi}_a, \eta^a)\dot{\eta}_a\dot{\eta}_b
$$
\begin{equation}\label{genericus}
+E^{a}(\phi_a, \dot{\phi}_a, \eta^a)\dot{\eta}_a+\lambda^a(\phi_a-F_a(\phi^b, \eta^a))+\beta^a(\eta_a-G_a(\phi^b, \eta^a)).
\end{equation}
This expression is a generalization of \eqref{cosa} to second order. It is  quite generic, except that the leading factor $A^{ab}$ is not allowed to depend on $\dot{\phi}_a$ and the dependence of the lagrangian on $\ddot{\phi}_a$ is quadratic and linear. The factor $D^{ab}$ is not allowed to depend on $\dot{\eta}_a$ and the dependence of the lagrangian on $\dot{\eta}_a$ is also quadratic and linear. Note that the fields $\phi_a$ involve up to second order derivatives, while the $\eta_a$ involve usual first order ones. That is the reason for doubling the notation, it is to indicate which fields contain higher order derivatives and which do not. 

The lagrange multipliers $\lambda^a$ and $\beta^a$ enforce that the fields may live in non trivial manifold. For instance, a sphere, or any other geometry. Here the dependence on $\phi_a$ may include spatial derivatives of these fields $\partial^i\phi_a$, the same follows for $\eta_a$. This is not written explicitly by notational simplicity.

Below, it will be assumed that the $\phi_a$ and the  above lagrangian are invariant under time reversal. This fixes the behavior of the unknown functions defining the action such as $A^{ab}$ and so on. 

In many situations, the above lagrangian involve apparent ghosts.  In those cases, the matrix $A^{ab}$ defines a negative quadratic form. This is true for the Pais-Uhlenbeck model, which contains apparent ghosts \cite{pais}. The following discussion relies on the method developed in \cite{salvio}-\cite{strumia3} for avoiding those ghosts.

Given  a higher order lagrangian such as the above one, it is convenient to enlarge the number of  variables by introducing  new ones $\psi_a$ by the relation
\begin{equation}\label{genericus2}
\psi_a=\alpha^{b}_a(\phi_b)\dot{\phi}_b,
\end{equation}
which represents $\dot{\phi}_a$ up to functions solely on $\phi_b$. The lagrangian has to be expressed in terms of these new variables. The Gauss-Ostrogradksy method for dealing with these theories in hamiltonian form consists of introducing canonically conjugated momentum densities to $\phi_a$ and $\psi_a$ given respectively by
$$
\pi^a= \frac{\partial L}{\partial \dot{\phi}_a}-\frac{d}{dt} \frac{\partial L}{\partial \ddot{\phi}_a},\qquad \Pi^a= \frac{\partial L}{\partial \dot{\psi}_a}-\frac{d}{dt} \frac{\partial L}{\partial \ddot{\psi}_a},\qquad P^a= \frac{\partial L}{\partial \dot{\eta}_a}.
$$
The standard classical Poisson brackets  lead to the same equations of motion than in the lagrangian formalism, for these higher order theories \cite{strumia1}-\cite{strumia3}.  The resulting hamiltonian is linear in $\pi^a$ and the Hamiltonian path integral formulation of the model is divergent. The references  \cite{salvio}-\cite{strumia3} propose to distinguish between the variables which are even under time reversal $Tq T^{-1}=q$ and those which are odd $Tq T^{-1}=-q$. For the first case one should implement usual quantization. For the second type of variables, known as the Dirac-Pauli type variables \cite{dirac}-\cite{pauli}, one has to employ a variant quantization.  

In the following, the result of the above procedure will be presented. The step by step details are collected in the appendix.

After applying this quantization procedure (see full details in the appendix)  the last expression becomes
$$
Z´=\int D(\alpha^b_a\dot{\phi}_b) D\eta_a D\Pi^c DP^a \exp\bigg\{\int [i\Pi^a\dot{\psi}_a+iP^a\dot{\eta}_a-H'(\psi^ a, \phi^b,\eta^a, \Pi^d, P^a)]d^4x\bigg\},
$$
or explicitly
$$
Z'=\int D(\alpha^b_a\dot{\phi}_b) D\eta_a D\Pi^c DP^a  \exp\bigg\{ \int [i\Pi^a\alpha^b_a\ddot{\phi}_b+i\Pi^a\frac{\delta \alpha^{c}_a}{\delta \phi_d}\dot{\phi}_c\dot{\phi}_d+iP^a\dot{\eta}_a
$$
$$
-\frac{1}{4}  A_{ab}\alpha^a_c\alpha^b_d(i\Pi^c+B^{m}(\phi_a,i\dot{\phi}_a,\eta^a)\gamma_m^c)(i\Pi^d-B^{m}(\phi_a,i\dot{\phi}_a,\eta^a)\gamma_m^d)
$$
$$
-\frac{1}{2}P^aD_{ab}(\phi_a,i\dot{\phi}_a,\eta^a)(P^b-E^b)
-i\Pi^a\frac{\delta \alpha^{c}_a}{\delta \phi_d}\dot{\phi}_c\dot{\phi}_d+C(\phi_a,i\dot{\phi}_a,\eta_a)
$$
$$
+\frac{1}{4}D_{cd}(\phi_a, i\dot{\phi}_a, \eta^a)(P^c-E^c(\phi_a, i\dot{\phi}_a, \eta^a))(P^d-E^d(\phi_a, i\dot{\phi}_a, \eta^a))
$$
$$
+\frac{1}{2}E^{a}(\phi_a, i\dot{\phi}_a, \eta^a)D_{ab}(\phi_a, i\dot{\phi}_a, \eta^a)(P^a+E^a(\phi_a, i\dot{\phi}_a, \eta^a))
$$
\begin{equation}\label{lugro2}
+\lambda^a(\phi_a-F_a(\phi^b, \eta^a))+\beta^a(\eta_a-G_a(\phi^b, \eta^a))]d^4x\bigg\}.
\end{equation}
All the intermediate steps leading to this formula are collected in the appendix. After simplifying some terms and performing the Gaussian integration on $\Pi^ a$ and $P^a$ by use of the Gaussian formula 
$$
\int d^nx \exp\{-\frac{1}{2}x^T U x+V^T x\} =\sqrt{\frac{(2\pi)^n}{\det U}}\exp\{\frac{1}{2}V^T U^{-1} V\},
$$
it is found 
$$
Z'=\int D(\alpha^b_a\dot{\phi}_b) D\eta_a\sqrt{\det (-A^{cd}\gamma_c^a \gamma_d^b)} a\sqrt{\det (D^{cd})}\exp\bigg\{ \int [A^{ab}(\phi_a,\eta_a)\ddot{\phi}_a \ddot{\phi}_b-B^{a}(\phi_a,i\dot{\phi}_a, \eta_a) \ddot{\phi}_b
$$
$$
+C(\phi_a,\eta_a,i\dot{\phi}_a)+D^{ab}(\phi_a, i\dot{\phi}_a, \eta^a)\dot{\eta}_a\dot{\eta}_b
+E^{a}(\phi_a, i\dot{\phi}_a, \eta^a)\dot{\eta}_a
$$
$$
+\lambda^a(\phi_a-F_a(\phi^b, \eta^a))+\beta^a(\eta_a-G_a(\phi^b, \eta^a))]d^4x\bigg\}.
$$
Going back from the euclidean  formulation change some signs and the final result becomes
$$
Z=\int D(\alpha^b_a\dot{\phi}_b) D\eta_a\sqrt{\det (-A^{cd}\gamma_c^a \gamma_d^b)} a\sqrt{\det (D^{cd})}\exp\bigg\{i \int [A^{ab}(\phi_a,\eta_a)\ddot{\phi}_a \ddot{\phi}_b+B^{a}(\phi_a,\dot{\phi}_a, \eta_a) \ddot{\phi}_b
$$
$$
+C(\phi_a,\eta_a,\dot{\phi}_a)+D^{ab}(\phi_a, \dot{\phi}_a, \eta^a)\dot{\eta}_a\dot{\eta}_b
+E^{a}(\phi_a, \dot{\phi}_a, \eta^a)\dot{\eta}_a
$$
$$
+\lambda^a(\phi_a-F_a(\phi^b, \eta^a))+\beta^a(\eta_a-G_a(\phi^b, \eta^a))]d^4x\bigg\}.
$$
In other words
$$
Z=\int D(\alpha^b_a\dot{\phi}_b) D\eta_a\sqrt{\det (-A^{cd}\gamma_c^a \gamma_d^b)} a\sqrt{\det (D^{cd})}e^{iS}.
$$
 The remaining step is to go from the integration in the variables $D(\alpha^b_a\dot{\phi}_b)$ to $D\phi_a$. The measure will be then
 $$
 d\mu_L=D(\phi_a) D\eta_a\sqrt{\det (-A^{cd}\gamma_c^a \gamma_d^b)} \sqrt{\det (D^{cd})}J,
 $$
 with $J$ the Jacobian for going from  $D(\alpha^b_a\dot{\phi}_b)$ to $D\phi_a$.  However, already in this form, the above expression will be useful for studying gravity. 
\\

\emph{Cancellations of the divergent terms proportional to $\delta^4(0)$.}
\\

Consider the simple situation in which $\alpha^a_b$ and consequently $\gamma_a^b$ are the identity matrix $\delta_a^b$. In this case the change of variables from $ D(\alpha^b_a\dot{\phi}_b)$ to $D\phi_a$ is direct, up to a factor proportional to $(\Delta t)^{-1}$ with $\Delta t$
the time interval defining the lattice in time, which in the continuum limit tends to zero.
 The measure in configuration space is given by
$$
d\mu_L(\phi_a)=\prod_{a,x} d\phi^a(x)\prod_{x,a} \sqrt{\det(-A^{ab}(\phi))}\sqrt{\det (D^{cd}(\phi))}.
$$
This measure can be expressed in exponential form as follows 
$$
d\mu_L(\phi_a)=\prod_{a,x} d\phi^a(x)\prod_{a} \exp\bigg\{\frac{1}{2} \delta^4(0)\int\log(\det(-A^{ab}(\phi)))d^4x\bigg\}
$$
\begin{equation}\label{mishure}
\exp\bigg\{\frac{1}{2} \delta^4(0)\int\log(\det(D^{ab}(\phi)))d^4x\bigg\}.
\end{equation}
This introduces an unwanted divergent factor proportional to $\delta^4(0)$, which one would like to cancel somehow. For quantization of gravity flat spaces the use of dimensional regularization ensures the Veltmann identity $\delta^4(0)=0$, but it is convenient to assume that this does not hold here if one is intended to study more general situations.

On the other hand, consider the action of the above model, expanded to second order in the fields around a classical solution $\phi_0^a(x)$ and $\eta_0^a(x)$
$$
S=S_0+\int d^4x\frac{\delta S}{\delta \phi^a(x)}\bigg|_{0}\delta \phi^a(x)
+\int d^4x\frac{\delta S}{\delta \eta^a(x)}\bigg|_{0}\delta \eta^a(x)
$$
$$
+\frac{1}{2}\int d^4x \int d^4y\frac{\delta^2 S}{\delta \phi^a(x)\delta \phi^b(y)}\bigg|_{0}\delta \phi^a(x)\delta \phi^b(y)+\frac{1}{2}\int d^4x \int d^4y\frac{\delta^2 S}{\delta \phi^a(x)\delta \eta^b(y)}\bigg|_{0}\delta \phi^a(x)\delta \eta^b(y)
$$
\begin{equation}\label{expandi}
+\frac{1}{2}\int d^4x \int d^4y\frac{\delta^2 S}{\delta \eta^a(x)\delta \phi^b(y)}\bigg|_{0}\delta \eta^a(x)\delta \phi^b(y)+\frac{1}{2}\int d^4x \int d^4y\frac{\delta^2 S}{\delta \eta^a(x)\delta \eta^b(y)}\bigg|_{0}\delta \eta^a(x)\delta \eta^b(y).
\end{equation}
The sign $|_0$ indicates that the functional derivatives are evaluated on the classical solutions. The second term vanishes, since  the classical equations of motion are satisfied. The last, for the theories under consideration \eqref{genericus}, can be worked out as
$$
\frac{\delta^2 S}{\delta \phi^a(x)\delta \phi^b(y)}\bigg|_{0}=[A^{ab}_{(0)}\partial_0^4+O_{3(0)}^{ab}\partial_0^3+O_{2(0)}^{ab}\partial_0^2
+O_{1(0)}^{ab}\partial_0+O_{0(0)}^{ab}]\delta^4(x-y).
$$
The explicit functional form of the terms $O^{ab}_{i(0)}=O^{ab}_i(\phi^a_0(x), \eta^a_0(x))$ with $i=0,..,3$ will not be relevant in the following discussion. These terms are related to the functions defining the lagrangian together with their derivatives on the fields, and may be simplified by assuming the validity of the equations of motion. The only important point in the following is that the higher order term is proportional to $A^{ab}(\phi_0^b(x))$ which is simultaneously the matrix defining the higher order term in the action.  In addition 
$$
\frac{\delta^2 S}{\delta \eta^a(x)\delta \eta^b(y)}\bigg|_{0}=[D_0^{ab}\partial_0^2+P_{2(0)}^{ab}\partial_0+P_{0(0)}^{ab}]\delta^4(x-y).
$$
The mixed variations $\frac{\delta^2 S}{\delta \phi^a(x)\delta \eta^b(y)}\bigg|_{0}$ and $\frac{\delta^2 S}{\delta \eta^a(x)\delta \phi^b(y)}\bigg|_{0}$ is at most of second order. The change of variable integrations in the path integral from $\phi^a(x)$ and $\eta^a$ to $\delta\phi^a(x)$ and $\delta\eta^a$, together with the last term, leads to a contribution to $Z$ 
$$
Z=e^{i S_0} Z_1,
$$
where $Z_1$ is related to the determinant of the inverse of the operator
\begin{equation}
 \hat{O}=  \left( \begin{array}{cc}
  \hat{\frac{\delta^2 S}{\delta \phi^a(x)\delta \phi^b(y)}}\bigg|_{0} & \hat{\frac{\delta^2 S}{\delta \phi^a(x)\delta \eta^b(y)}}\bigg|_{0} \\
 \hat{\frac{\delta^2 S}{\delta \eta^a(x)\delta \phi^b(y)}}\bigg|_{0} &\hat{\frac{\delta^2 S}{\delta \eta^a(x)\delta \eta^b(y)}}\bigg|_{0}
\end{array}\right) =\left( \begin{array}{cc}
  U & V \\
W &Z
\end{array}\right),
\end{equation}
the symbol $\hat{}$  in the matrix elements denotes the action of removing of the factor $\delta^4(x-y)$ in the corresponding expression. The last formula itself gives the definition of the operators $U$, $V$, $W$ and $Z$. Employ now the Schur formula for determinants
$$
\det(O)=\det(U)\det(Z-W U^{-1}V).
$$
This leads to two contributions to the determinant. Consider first the one related to $U$. The corresponding fluctuation part has the following form
$$
-\log Z^U_1=\frac{1}{2}\text{Tr} \log[A^{ab}_{(0)}\partial_0^4+O_{3(0)}^{ab}\partial_0^3+O_{2(0)}^{ab}\partial_0^2
+O_{1(0)}^{ab}\partial_0+O_{0(0)}^{ab})]].
$$
Here the trace of a given operator $O$ can be calculated as the sum of its eigenvalues, or by the formula
$$
\text{Tr}\,O=\int d^4x \lim_{y\to x}O\,\delta^4(x-y),
$$
if the operator is suitable enough. In fact, if one can find a set of eigenfunctions $\psi_n(x)$ of $O$ with eigenvalues $\lambda_n$, allowing to represent the delta function by the completeness relation
$$
\delta^4(x-y)=\sum_n \psi_n(x)\psi_n^\ast(y), 
$$
this,  when inserted into the last expression, leads to the standard notion of a trace
$$
\text{Tr}\, O=\sum _n\lambda_n,
$$
by assuming that the eigenfunctions $\psi_n(x)$ constitute  an orthonormal basis. Going back to the fluctuation contribution to the the path integral given in $Z^U_1$, it is convenient to use $\log(ab)=\log a+\log b$  in order  to  pick the higher order operator in the following way
$$
-\log Z^U_1=\frac{1}{2}\text{Tr} \log[A^{ab}_{(0)})\partial_0^4]+\frac{1}{2}\text{Tr} \log[1+A_{(0)ac}O_{3(0)}^{cb}\partial^{-1}_0+A_{(0)ab} O_{2(0)}^{cb}\partial_0^{-2}
$$
$$
+A_{(0)ac}O_{1(0)}^{cb}\partial^{-3}_0+A_{ac(0)}O_{0(0)}^{cb}\partial_0^{-4}]].
$$
 The second logarithm may be expanded in Taylor series around $1$  leading to 
$$
-\log Z^U_1=\frac{1}{2}\text{Tr} \log[A^{ab}_{(0)}\partial_0^4]+\frac{1}{2}\text{Tr}\sum_n \frac{1}{n}[A_{ac(0)}O_{3(0)}^{cb}\partial^{-1}_0+A_{ac(0)}O_{2(0)}^{cb}\partial_0^{-2}
$$
$$
+A_{ac(0)}O_{1(0)}^{cb}\partial^{-3}_0+A_{ac(0)}O_{0(0)}^{cb}\partial_0^{-4}]^n.
$$
However, the most important contribution comes from the first term, which still retains the logarithm, and it can be expressed as 
$$
\frac{1}{2}\text{Tr} \log[A^{ab}_{(0)}\partial_0^4]=\frac{1}{2}\int d^4x \lim_{y\to x}\log[A^{ab}_{(0)}\partial_0^4]\,\delta^4(x-y)
$$
$$=\frac{1}{2}\int d^4x \lim_{y\to x}\log  A^{ab}_{(0)}\,\delta^4(x-y)+ \frac{1}{2}\text{Tr}\log\partial_0^4
$$
$$
=\frac{1}{2}\delta^4(0)\text{Tr}\int d^4x \log  A^{ab}_{(0)}+ \frac{1}{2}\text{Tr}\log\partial_0^4.
$$
The last traces are only related to the indices $a$ and $b$.  The last formula shows that generically, a non trivial matrix $ A^{ab}_{(0)}$ leads to a term proportional to $\delta^4(0)$. However, this term precisely cancels the one appearing  in the measure \eqref{mishure} related to $A^{ab}_{(0)}$. Therefore, the above calculation generalize the features found in \cite{fradkin} for quartic theories of the type \eqref{mia}, on the extremal.

Note that the presence of the term indicated above is induced by the elementary property of the logarithm of a product, which separates the term proportional to $ A^{ab}_{(0)}$ . The remaining terms do not contain a logarithm and this separation does not take place. They alone do not give further $\delta^4(0)$ contributions. The only issue that may appear is that the full sum is divergent, however this can be taken care with other regularization methods.

It remains to study the contribution to the determinant of the term $Z-W U^{-1} V$. As $U$ is of fourth order
and $W$ and $V$ are of second order, the leading term is the higher order one in $Z$, which is proportional to $D^{ab}_{(0)}$. Repetition of the above arguments will show the presence of a term $\frac{1}{2}\delta^4(0)\text{Tr}\int d^4x \log  D^{ab}_{(0)}$ canceling the divergent part of the measure \eqref{mishure} related to $D^{ab}_{(0)}$.

The reader is challenged to extend the above reasoning  to theories with derivative orders higher than four.

\section{Quadratic gravity}
\subsection{Synchronous gauge}
Even the simple case with $\alpha_a^b=c\;\delta^a_b$, with $c$ a constant is useful for studying the cancellations of the terms $\delta^4(0)$ for  Stelle gravity \cite{stelle1}-\cite{stelle2}, whose action is given in \eqref{geld}. In the synchronous gauge in which the metric is expressed as 
$$
ds^2 = -dt^2+g_{ij}(x)dx^idx^j,
$$
it will be shown below that the Stelle model belongs to the type of theories just discussed above. This follows from the fact that,  after defining the Dirac-Pauli variable
\begin{equation}\label{abovedefi}
K_{ij}=\frac{1}{2}\dot{g}_{ij},
\end{equation}
the lagrangian \eqref{geld} takes the following form
$$
L=-\kappa^{-2}(- 2 g^{ij} \dot K_{ij}-3K_{ij}K^{ij} R^{(3)}+K^2)+\beta(- 2 g^{ij} \dot K_{ij}-3K_{ij}K^{ij} R^{(3)}+K^2)^2
$$
$$
+\alpha \bigg\{\left[g^{ij} (\dot K_{ij}+K_{il}K^l_{~j})\right]^2-2(\nabla^{(3)}_i K-\nabla^{(3)j}K_{ji})(\nabla^{(3)i}K-\nabla^{(3)l}K_l^{~i}) 
$$
$$
+(~R^{(3)}_{ij}-\dot K_{ij}-2K_{il}K^l_{~j}+KK_{ij})g^{il}g^{jm}(~R^{(3)}_{lm}-\dot K_{lm}-2K_{lp}K^p_{~m}+KK_{lm})\bigg\}.
$$
A simple inspection shows that this lagrangian  is of the form \eqref{genericus}.  To see this, note that the above definition \eqref{abovedefi} of $K_{ij}$ can be seen to be a particular case of  \eqref{genericus2} with constant $\alpha^a_b$ matrices.   In additon, it is helpful to identify $\phi_a$ with $g_{ij}$ and $K_{ij}$ with $\psi_a$ and to notice that the quadratic terms in $\dot{K}_{ij}$ are multiplied by coefficients that depend only on the metric tensor $g^{ij}$, not on $K_{ij}$ itself.  The synchronous gauge can be enforced with the introduction of lagrange multipliers $\lambda_a$ ensuring that $g_{44}-1=g_{i4}=0$ with $i=1,2,3$. The Faddeev Popov determinant in this case is unity, therefore no superdeterminants \cite{dewito} are involved. The discussion of the previous subsection can be repeated fully for this case, and shows the cancellation of the divergences $\delta^4(0)$ in the extremal for the synchronous gauge in Stelle gravity.

\subsection{Other gauges and superdeterminants}

It is natural at this point to ask if the cancellation of volume divergences is true for non synchronous gauges, in which the metric is decomposed as
\begin{equation}
 g_{\mu\nu} = 
 \left( \begin{array}{cc}
  - N^2 + N_i N^i & N_i \\
  N_i & g_{ij}
\end{array}\right) \,.
\end{equation}
Introduce the additional variable
\begin{equation}
	K_{ij} = \frac{1}{2N} \left( \dot{g}_{ij} - 2 \nabla_{(i} N_{j)} \right) \,.
\end{equation}
Note that this is a Dirac-Pauli variable as well, since the $N_i$ correspond to the components of the metric $g_{i0}$ which are odd under time reversal. Furthermore $N$ is related to $g_{00}$ and is even under time reversal. The Stelle lagrangian \eqref{geld} becomes in this choice of coordinates expressed as follows \cite{chile}-\cite{chile2}
$$
L=  \sqrt{g} N [ - \kappa^{-2} \left( 2 E + R + L \right)
 + \alpha \left( ( E_{ij} + R_{ij} ) ( E^{ij} +  R^{ij} \right) 
$$
$$
+ ( E + L )^2 - 2 V_i V^i+ \beta \left( 2 E + R + L \right)^2 \Big] \,,
$$
where now the curvatures $R_{ij}$ are constructed with the spatial components $g_{ij}$ and depend on time $t$ as parameter, their indices are raised and lowered with $g_{ij}$ and its inverse $g^{ij}$ and the following quantities 
$$
	E_{ij} = \frac{1}{N} \left( \dot{K}_{ij} - \nabla_{ij} N 
	- N ( 2 K_{ik} K_{j}{}^{k} - K K_{ij} )
	- 2 K_{k(i} \nabla_{j)} N^k - N^k \nabla_k K_{ij} \right) ,
	$$$$
	L = K_{ij} K^{ij} - K^2 \qquad
	V_i = \nabla^j K_{ij} - \nabla_i K ,
$$
have been introduced. The functions $N_i$ do not have time derivatives, therefore they lead to constraints. The above lagrangian can be seen, after some inspection, to depend on $\dot{K}_{ij}$ with quadratic and linear terms, the quadratic only depend on $g_{ij}$, $N$ and $N_i$ without their time derivatives. This imitates the features described in \eqref{mia}.

The time derivative of the metric can be expressed in terms of $K_{ij}$ as
$$\dot{g}_{ij} = 2 N K_{ij} + 2 \nabla_{(i} N_{j)}.$$
This corresponds to a choice of canonical coordinates  $(g_{ij},\pi^{ij}), (K_{ij},P^{ij}), (N,P_N), (N^i,P_i)$. In particular, 
\begin{equation}
P^{ij} = \frac{ \delta \mathcal{L} }{ \delta \dot{K}_{ij} } =
\sqrt{g} G^{ijkl} E_{kl} 
+ 2 \sqrt{g} \left( \alpha  R^{ij} - g^{ij} X \right) \,. 
\end{equation}
where $X = \kappa^{-2} - \upsilon_2 L - 2 \beta R$ with $v_2=\alpha+2\beta$  and
$$  
 G^{ijkl} =  
   \alpha ( g^{ik} g^{jl} + g^{il} g^{jk} )
 + 2 \upsilon_4 g^{ij} g^{kl} ,
 $$
$$
{G}^{-1}_{ijkl} =
   (4\alpha)^{-1} ( g_{ik} g_{jl} + g_{il} g_{jk} )
 - \gamma\upsilon_4 g_{ij} g_{kl},
 $$
$$
G^{i j k l} G^{-1}_{k l m n} = \frac{1}{2} ( \delta^i_m \delta^j_n + \delta^i_n \delta^j_m ).
$$
The time derivatives of $\dot{K}_{ij}$ can be obtained by the above expression for $P^{ij}$, the result is
\begin{equation}
\dot{K}_{ij} = 
\frac{N}{\sqrt{g}} \tilde{Q}_{i j} + B_{ij} .
\end{equation}
where
$$
 Q^{i j} = 
 P^{i j} - 2 \sqrt{g} \left( \alpha  R^{ij} - g^{ij} X \right)  
 $$
 $$ 
 \tilde{Q}_{ij} = G^{-1}_{ijkl} Q^{kl},
 \qquad
 \hat{Q}_{ij} = \tilde{Q}_{ij} + \sqrt{g} R_{ij} \,,
 $$
  $$
J_{ijkl} = 2 K_{ik} K_{jl} - K_{ij} K_{kl} \,,
 \quad J_{ij} = g^{kl} J_{ijkl}.$$
$$
 B_{ij} = 
 \left( \nabla_{i}\nabla_j + J_{ij} \right) N
 + 2 K_{k(i} \nabla_{j)} N^k + N^k \nabla_k K_{ij} 
  $$
The variables $N$ and $N_i$ are not dynamical. This leads to several constraints, primary and secondary, and first and of second order \cite{buchbinder}-\cite{buchbinder2}.  These constraints were recently studied in great detail in \cite{chile}-\cite{chile2}. After quantization by use of constrained system techniques, the resulting path integral becomes
$$
Z=\int\prod_{x} dg_{\mu\nu}(x) (g^{00}(x))^4 |g(x)|^{\frac{3}{2}} \Delta_{f} e^{iS}
$$
where $\Delta_f$ is the Faddeev-Popov determinant related to the chosen gauge fixing.

The role of the determinant $\Delta_f$ is not to be overlooked. Faddeev-Popov determinants are responsible for the presence of ghost lagrangian to the action. These ghosts are  the cause of several important physical effects, for example anomalies or to additional contributions the value of central charges in a given conformal field theory. In general, ghost fields have kinetic terms and  are coupled to the metric tensor $g_{\mu\nu}$. In order to give an example, in a Verlinde-Verlinde gauge \cite{Verlinde} 
\begin{equation}
 g^4_{\mu\nu} = 
 \left( \begin{array}{cc}
  g_{\alpha\beta} & 0 \\
  0 & h_{ij}
\end{array}\right) ,
\end{equation}
the ghost action is
\begin{equation}\label{slide}
S_g=\int \sqrt{g}\sqrt{h} [h^{ij}b_{i\alpha}\partial_j c^{\alpha}+g^{\alpha\beta}b_{\alpha j}\partial_\beta c^{j}]d4x.
\end{equation}
This  example shows that the ghosts $c^\mu$ and $b_{\mu\nu}$ are generically coupled to the metric components $g_{\mu\nu}$. When studying fluctuations on $Z$ around an extremal, one may consider the effect of those which involve not only the graviton, but ghosts as well. These extremals may induce $\delta^4(0)$ terms which involve the functions defining the kinetic terms of the ghosts.

As far as i understand, this motivates the reference \cite{unz} to consider superdeterminants \cite{dewito}.  If one is working with expansions around extremal solutions with the ghosts $b_{\mu\nu}$ and $c^\mu$ turned off, the measure of  \cite{buchbinder}-\cite{buchbinder2} are enough for compensating the volume divergences, by the results presented in the previous section. Assume that these fields are turned on instead, and consider a lagrangian of the form \eqref{genericus} but involving usual and Grassmann variables. If $A^{ab}$ and $D^{ab}$ are non degenerate then the resulting determinants $\det{A^{ab}}$ and $\det D^{ab}$ should be replaced by  s$\det{A^{ab}}$ and s$\det D^{ab}$, where s$\det$ is the super-determinant. Therefore, the most effective way to generate the measure may be to start with the lagrangian of Quadratic  Gravity with a given Faddeev-Popov action, and then employ hamiltonian quantization of the full gravity and ghost system. The integration of the 
momentum will lead to a measure which will contain a superdeterminant, since the resulting matrix will generically  contain  usual and Grassmann variables. The cancellation of volume divergences $\delta^4(0)$ in the extremal is ensured, under the assumption that  the ghost lagrangian is chosen in such a way that the lagrangian form \eqref{genericus} applies.

Note that there may be general gauge choices in which the form \eqref{genericus} does not take place. The discussion given above is not warranted to apply in those cases.

\section{Discussion}

In the present work it was shown that the measure \cite{buchbinder}-\cite{buchbinder2} for Quadratic Gravity has been shown to ensure cancellation of the volume divergences $\delta^4(0)$, at least in the extremal. This is up to subtle details related to superdeterminants, discussed below formula \eqref{slide}. This imitates the features found in \cite{fradkin} for their measure in GR.

The above result alone of course, does not imply that these measures are the correct ones. They only imply that these measures posses in the extremal a nice property which is usual in dimensional regularization namely, the vanishing of the terms proportional to $\delta^4(0)$.  This fact is not mandatory, it comes as a theoretical bonus. Note that there exist references such as \cite{anselmi}-\cite{anselmi2} which consider the possibility of propagating these divergences and dealing with them with counter terms. This approach is perfectly legitimate, however, it may be hard to apply.  

For this reason, other possible invariant measures for Quantum Gravity were considered. The first type of measure studied are invariant under coordinate transformations, and mimics the properties considered in \cite{botelho}, \cite{mottola1}-\cite{mottola5} based on the earlier work \cite{dewitt}.  In this formalism, the fact that the measure is invariant under general coordinate transformations is justified without relying on the gravity model under study. The approach is geometric and introduces a natural notion of distance in the space of metric component configurations leading to the invariant measure. These types of measures can be applied to almost any quantum theory of gravity, since its invariance under coordinate transformations can be justified by its purely geometric arguments.  In four dimensions, the simplest choice was derived in \eqref{mia}. However, in these cases, there is a propagating $\delta^4(0)$ term in the action which complicates the renormalization program. Of course, when quantization along flat space is done, and dimensional regularization is employed, then the Veltman identities set these quantities to zero. In curved spaces, for higher loops, these identities or some alternative covariant method of quantization allowing the avoidance of volume divergences has to be worked out independently.

It is important to remark that formally, there is no reason for requesting that the volume divergences cancel out. One may work a renormalization program which takes care of them independently. However, this may be hard.

A crucial difference between the measures considered by the authors \cite{botelho}, \cite{mottola1}-\cite{mottola5} and \cite{dewitt} with the ones found in \cite{fradkin}, \cite{buchbinder}-\cite{buchbinder2}, is that the last ones strongly depend on the model under study.  The invariance of these measures is justified by assuming that the Feynman path integral $Z$ is invariant under different choices of canonical transformations. This is applied in order to determine the measure and its change under general coordinate transformation, by assuming that in all the intermediate steps the action is given by a generic type of lagrangian such as \eqref{cosa}. All the steps involved are dependent on the lagrangian in consideration, and so does the measure. The result contains factors in powers of $g^{00}$. The argument however may be dubious if there appears an anomaly invalidating the general form \eqref{cosa}.

These measures have been source of debate and their use in practical applications was avoided in several references \cite{kevin}-\cite{kevin4}. While this approach is grounded on concrete calculations and can not be criticized, the authors opinion is that  the arguments so far do not constitute an elimination point. There are works which instead, prefer the use of the non flat measure \cite{kuntz1}-\cite{kuntz6}.  These apparently non covariant measures can be discarded only if it is proved that their anomaly under general coordinate transformations can not be adsorbed with the allowed counterterms of the model. Since the full counterterm structure of Stelle gravity, perturbed around curved backgrounds is not yet established, it is difficult to reject them.  

Note that the method \cite{fradkin}, when applied to an scalar fields, give rise to an anomalous measure. However, this anomalous measure induce counter term redefinitions and is perfectly acceptable \cite{nieu}. It would be a nice feature if this was the case in gravity, suggesting the avoidance of the divergent $\delta^4(0)$  terms from the very beginning.
The same follows for the measures of \cite{buchbinder}-\cite{buchbinder2}. It is perhaps worthy to consider general measures depending on powers of $g^a$ and $(g^{00})^b$, and to analyze the propagation of these factors in specific calculations, instead of throwing them away from the very beginning.

It should be emphasized that the seminal  works \cite{hawking}, \cite{seminal} about heat kernel regularization also allow to overcome the problem of the $\delta^4(0)$ in curved backgrounds. When applying this technique, the precise determination of the measure is not relevant.  These results however hold at one loop. There are attempts to generalize these methods to higher loop corrections, for instance de Schwinger DeWitt method  or the heat kernel method. Reviews can be found in  \cite{hit1}-\cite{hit6}. Generalizations of the zeta regularization to multiloops are considered in \cite{bilal}. These techniques are of obvious interest and may be an alternative of the difficult problem of choosing the measure. These developments of course deserve further attention.

\section*{Acknowledgements}
The author is supported by CONICET, Argentina and by the Grant
PICT 2020-02181.

\appendix

\section{Some explicit calculations}
In this appendix, the explicit pass from the lagrangian \eqref{genericus} to the expression \eqref{lugro2} will be derived.
From this lagrangian and \eqref{genericus2} it is seen that
\begin{equation}\label{aside}
\dot{\phi}_a=\gamma^b_a(\phi_c) \psi_b,
\end{equation}
with $\gamma^a_b$ the inverse of $\alpha^a_b$
$$
\gamma^a_b\alpha^b_c=\delta_c^b.
$$
The time derivatives of the last expressions can be worked out to give
$$
\dot{\psi}_a=\alpha^{b}_a(\phi_b)\ddot{\phi}_b+\frac{\delta \alpha^{b}_a}{\delta \phi_c}\dot{\phi}_b\dot{\phi}_c, 
$$
and consequently
$$
\ddot{\phi}_a=\gamma^{b}_a[\dot{\psi}_b-\frac{\delta \alpha^{c}_b}{\delta \phi_d}\dot{\phi}_c\dot{\phi}_d].
$$
The lagrangian in the new variables is expressed as follows
$$
L=A^{ab}(\phi_a)\gamma^{e}_a[\dot{\psi}_e-\frac{\delta \alpha^{c}_e}{\delta \phi_d}\dot{\phi}_c\dot{\phi}_d]\gamma^{f}_b[\dot{\psi}_f-\frac{\delta \alpha^{c}_f}{\delta \phi_d}\dot{\phi}_c\dot{\phi}_d]
$$
$$
+B^{a}(\phi_a,\dot{\phi}_a) \gamma^{e}_a[\dot{\psi}_e-\frac{\delta \alpha^{c}_e}{\delta \phi_d}\dot{\phi}_c\dot{\phi}_d]+C(\phi_a,\dot{\phi}_a)+D^{ab}(\phi_a, \dot{\phi}_a, \eta^a)\dot{\eta}_a\dot{\eta}_b
$$
$$
+E^{a}(\phi_a, \dot{\phi}_a, \eta^a)\dot{\eta}_a+\lambda^a(\phi_a-F_a(\phi^b, \eta^a))+\beta^a(\eta_a-G_a(\phi^b, \eta^a)).
$$
The Gauss-Ostrogradksy method for dealing with these theories in hamiltonian form consists of introducing canonically conjugated momentum densities to $\phi_a$ and $\psi_a$ given respectively by
$$
\pi^a= \frac{\partial L}{\partial \dot{\phi}_a}-\frac{d}{dt} \frac{\partial L}{\partial \ddot{\phi}_a},\qquad \Pi^a= \frac{\partial L}{\partial \dot{\psi}_a}-\frac{d}{dt} \frac{\partial L}{\partial \ddot{\psi}_a},\qquad P^a= \frac{\partial L}{\partial \dot{\eta}_a}.
$$
The standard classical Poisson brackets  lead to the same equations of motion than in the lagrangian formalism, for these higher order theories \cite{strumia1}-\cite{strumia3}. 

The momentum $P^a$ related to the standard variable $\eta^a$ is calculated easily  
$$
P^a=2D^{ab}(\phi_a, \dot{\phi}_a, \eta^a)\dot{\eta}_a
+E^{a}(\phi_a, \dot{\phi}_a, \eta^a), 
$$
and can be inverted in order to express  $\dot{\eta}_a$ as
$$
\dot{\eta}_a=\frac{1}{2}D_{ab}(P^b-E^b),
$$
with $D_{ab}$ the inverse of $D^{ab}$. The second formula defining $\Pi^a$ leads to
$$
\Pi^a=2A^{eb}(\phi_a)\gamma^{a}_e\gamma^{f}_b[\dot{\psi}_f-\frac{\delta \alpha^{c}_f}{\delta \phi_d}\dot{\phi}_c\dot{\phi}_d]+B^{c} \gamma^{a}_c,
$$
the calculation is simple since the  lagragian does not depend on $\ddot{\psi}_a$.  The last formula can be inverted in order to express the derivative $\dot{\psi}_a$ as
$$
\dot{\psi}_a=\frac{1}{2} A_{mq}\alpha^m_a \alpha^q_c(\Pi^c-B^{n}\gamma_n^c)+\frac{\delta \alpha^{c}_a}{\delta \phi_d}\dot{\phi}_c\dot{\phi}_d,
$$
with $A_{ab}(\phi)$ the inverse of $A^{ab}(\phi)$.  The explicit expression for $\pi^a$ is more involved, however it will not be relevant in the following, and is omitted for this reason.

The lagrangian in the new variables is
$$
L=\frac{1}{4}  A_{ab}\alpha^a_c\alpha^b_d(\Pi^c-B^{m}\gamma_m^c)(\Pi^d-B^{m}\gamma_m^d)
+\frac{1}{2} B^{a}(\phi_a,\dot{\phi}_a) A_{aq}\alpha^q_c(\Pi^c-B^{m}\gamma_m^c)
$$
$$
+C(\phi_a,\dot{\phi}_a,\eta_a)+\frac{1}{4}{}D^{ab}(\phi_a, \dot{\phi}_a, \eta^a)D_{ac}(P^c-E^c)D_{bd}(P^d-E^d)+\frac{1}{2}E^{a}(\phi_a, \dot{\phi}_a, \eta^a)D_{ab}(P^a-E^a)
$$
$$
+\lambda^a(\phi_a-F_a(\phi^b, \eta^a))+\beta^a(\eta_a-G_a(\phi^b, \eta^a)),
$$
and can be written  by use of \eqref{aside} as a function $L(\Pi^a P^a,\phi_a, \psi_a, \eta_a)$ without time derivatives. It does not depend on $\pi^a$ either, that is the reason for which the calculation of this quantity was not presented above.

The hamiltonian density of the system
$$
H=\pi^a\dot{\phi}_a+\Pi^a\dot{\psi}_a+P^a\dot{\eta}_a-L,
$$
after employing the above replacements becomes
$$
H=\pi^a\gamma^b_a(\phi_c) \psi_b+\frac{1}{4}  A_{ab}\alpha^a_c\alpha^b_d(\Pi^c-B^{m}\gamma_m^c)(\Pi^d-B^{m}\gamma_m^d)+\frac{1}{2}P^aD_{ab}(P^b-E^b)
$$
$$
+\Pi^a\frac{\delta \alpha^{c}_a}{\delta \phi_d}\dot{\phi}_c\dot{\phi}_d-C(\phi_a,\dot{\phi}_a,\eta_a)
-\frac{1}{4}{}D^{ab}(\phi_a, \dot{\phi}_a, \eta^a)D_{ac}(P^c-E^c)D_{bd}(P^d-E^d)
$$
$$
-\frac{1}{2}E^{a}(\phi_a, \dot{\phi}_a, \eta^a)D_{ab}(P^a-E^a)
-\lambda^a(\phi_a-F_a(\phi^b, \eta^a))-\beta^a(\eta_a-G_a(\phi^b, \eta^a)).
$$
The quantity $\pi^a$ appears in a linear term. Therefore, if the euclidean path integral is employed, this term makes the integration unbounded.  This type of situation appears already in the simplest of these models, the Pais-Uhlenbeck model \cite{pais}.  A way to circumvent this ghost problem was found in \cite{salvio}-\cite{strumia3}.  In this reference, it is proposed to distinguish between the variables which are even under time reversal $Tq T^{-1}=q$ and those which are odd $Tq T^{-1}=-q$. For the first case one should implement usual quantization. For the second type of variables, known as the Dirac-Pauli type variables \cite{dirac}-\cite{pauli}, one has to employ a variant quantization.  This variant requires to employ the path integral in euclidean Hamitonian form which, for the Liouville choice of the measure,  is given by the expression
$$
Z'=\int D\phi_a  D\psi_b D\eta^a D\Pi^c D\pi^d dP^a\exp\bigg\{ \int [i\pi^a\dot{\phi}_a+i\Pi^a\dot{\psi}_a+iP^a \eta^a-H'(\psi^ a, \phi^b, \eta^a,\pi^c, \Pi^d, P^a)]d^4x\bigg\}.
$$
Here the hamiltonian $H'$ corresponds to the original hamiltonian in which the Dirac-Pauli variables are 
redefined as $q_{DP}\to i q_{DP}$, and their corresponding impulses are transformed as $p_{DP}\to -ip_{DP}$. The time is the euclidean one $\tau$ and the result has to be analitically continued to $\tau\to  it$. In the present model $\dot{\phi}_a$  is  a Dirac-Pauli variable, so is $\psi_a$. The quantization procedure of this references implies to make the replacement in the hamiltonian $\psi_a\to i\psi_a$ and $\Pi^a\to -i\Pi^a$. This results in the following hamiltonian 
$$
H´=\pi^a\gamma^b_a(\phi_c) i\psi_b+\frac{1}{4}  A_{ab}\alpha^a_c \alpha^b_d(i\Pi^c+B^m(\phi_a, i\dot{\phi}_a,\eta^a)\gamma_m^c)(i\Pi^d+B^m(\phi_a, i\dot{\phi}_a, \eta_a)\gamma_m^d)
$$
$$
+\frac{1}{2}P^aD_{ab}(\phi_a,i\dot{\phi}_a,\eta^a)(P^b-E^b)
+i\Pi^a\frac{\delta \alpha^{c}_a}{\delta \phi_d}\dot{\phi}_c\dot{\phi}_d-C(\phi_a,i\dot{\phi}_a,\eta_a)
$$
$$
-\frac{1}{4}D^{ab}(\phi_a, i\dot{\phi}_a, \eta^a)D_{ac}(\phi_a, i\dot{\phi}_a, \eta^a)(P^c-E^c(\phi_a, i\dot{\phi}_a, \eta^a))D_{bd}(\phi_a, i\dot{\phi}_a, \eta^a)(P^d-E^d(\phi_a, i\dot{\phi}_a, \eta^a))
$$
$$
-\frac{1}{2}E^{a}(\phi_a, i\dot{\phi}_a, \eta^a)D_{ab}(\phi_a, i\dot{\phi}_a, \eta^a)(P^a-E^a(\phi_a, i\dot{\phi}_a, \eta^a))
-\lambda^a(\phi_a-F_a(\phi^b, \eta^a))-\beta^a(\eta_a-G_a(\phi^b, \eta^a)).
$$
In the last expression it is assumed that $\dot{\phi}_a$ is expressed in terms of $\psi_a$ by use of \eqref{aside}, but this is not written explicitly in order to avoid cumbersome expressions. Then, the integration in $\pi^a$ in $Z´$ gives a functional Dirac delta $\delta(\psi_a-\alpha^b_a\dot{\phi}_b)$ fixing $\psi_a=\alpha^b_a\dot{\phi}_b$. Therefore, the last expression becomes
$$
Z´=\int D(\alpha^b_a\dot{\phi}_b) D\eta_a D\Pi^c DP^a \exp\bigg\{\int [i\Pi^a\dot{\psi}_a+iP^a\dot{\eta}_a-H'(\psi^ a, \phi^b,\eta^a, \Pi^d, P^a)]d^4x\bigg\},
$$
or explicitly
$$
Z'=\int D(\alpha^b_a\dot{\phi}_b) D\eta_a D\Pi^c DP^a  \exp\bigg\{ \int [i\Pi^a\alpha^b_a\ddot{\phi}_b+i\Pi^a\frac{\delta \alpha^{c}_a}{\delta \phi_d}\dot{\phi}_c\dot{\phi}_d+iP^a\dot{\eta}_a
$$
$$
-\frac{1}{4}  A_{ab}\alpha^a_c\alpha^b_d(i\Pi^c+B^{m}(\phi_a,i\dot{\phi}_a,\eta^a)\gamma_m^c)(i\Pi^d-B^{m}(\phi_a,i\dot{\phi}_a,\eta^a)\gamma_m^d)
$$
$$
-\frac{1}{2}P^aD_{ab}(\phi_a,i\dot{\phi}_a,\eta^a)(P^b-E^b)
-i\Pi^a\frac{\delta \alpha^{c}_a}{\delta \phi_d}\dot{\phi}_c\dot{\phi}_d+C(\phi_a,i\dot{\phi}_a,\eta_a)
$$
$$
+\frac{1}{4}D_{cd}(\phi_a, i\dot{\phi}_a, \eta^a)(P^c-E^c(\phi_a, i\dot{\phi}_a, \eta^a))(P^d-E^d(\phi_a, i\dot{\phi}_a, \eta^a))
$$
$$
+\frac{1}{2}E^{a}(\phi_a, i\dot{\phi}_a, \eta^a)D_{ab}(\phi_a, i\dot{\phi}_a, \eta^a)(P^a+E^a(\phi_a, i\dot{\phi}_a, \eta^a))
$$
$$
+\lambda^a(\phi_a-F_a(\phi^b, \eta^a))+\beta^a(\eta_a-G_a(\phi^b, \eta^a))]d^4x\bigg\}.
$$
This is exactly \eqref{lugro2}.

\end{document}